\documentclass[a4paper,12pt]{article}  

\usepackage{color}
\usepackage{amssymb}
\usepackage{amsmath}
\usepackage{graphicx}
\usepackage{authblk}
\usepackage{leftidx}
\usepackage{footmisc}
\usepackage{mathrsfs}

\usepackage{hyperref}
\hypersetup{
   colorlinks=true,         
   linkcolor=red,          
   citecolor=red,        
   urlcolor=red            
}

\usepackage{stmaryrd}

\newcommand{\Supnorm}[1]{{\color{black}\|}#1{\color{black}\|_{L^\infty}}}

\title{What is a reduced boundary in general relativity?}

\date{\today}

\author{Emmanuele Battista$^{1,2,*}$ ORCID: 0000-0001-5361-7109}
\author{Giampiero Esposito$^{3,\dagger}$ ORCID: 0000-0001-5930-8366}
\affil{$^1$ Institute for Theoretical Physics, Karlsruhe Institute of Technology (KIT), 
76128 Karlsruhe, Germany\\
$^2$ Institute for Nuclear Physics, Karlsruhe Institute of Technology (KIT), 
Hermann-von-Helmholtz-Platz 1, 76344 Eggenstein-Leopoldshafen, Germany\\
$^3$ Istituto Nazionale di Fisica Nucleare, Sezione di Napoli, Complesso Universitario 
di Monte S. Angelo, Via Cintia Edificio 6, 80126 Napoli, Italy\\
$^*$ emmanuele.battista@kit.edu\\
$^\dagger$ gesposit@na.infn.it}

\begin{document}
\maketitle

\begin{abstract}
Th\-e con\-ce\-pt of bo\-un\-da\-ry pla\-ys an im\-por\-ta\-nt ro\-le 
in se\-ve\-ral bran\-ches of ge\-ne\-ral re\-la\-ti\-vi\-ty,
e.g., the variational principle for the Einstein equations, the event horizon and the 
apparent horizon of black holes, the formation of trapped surfaces. 
On the other hand, in a branch of mathematics known as geometric
measure theory, the usefulness has been discovered long ago of 
yet another concept, i.e., the reduced
boundary of a finite-perimeter set. This paper proposes therefore a definition 
of finite-perimeter sets and their reduced boundary in general relativity.
Moreover, a basic integral formula of geometric measure theory is evaluated explicitly
in the relevant case of Euclidean Schwarzschild geometry,
for the first time in the literature. This research prepares 
the ground for a measure-theoretic approach to several concepts
in gravitational physics, supplemented by geometric insight. 
Moreover, such an investigation suggests considering the possibility that
the in-out amplitude for Euclidean quantum gravity should be evaluated over
finite-perimeter Riemannian geometries that match the assigned data
on their reduced boundary. As a possible application, an analysis is 
performed of the basic formulae leading eventually to the corrections of the 
intrinsic quantum mechanical entropy of a black hole. 
\end{abstract}

\section{Introduction}
\setcounter{equation}{0}

When Fermi was investigating the nature of mesons with his student Yang in the late forties, 
they conceived a title of their paper with a question mark at the end \cite{1949}.
In the course of completing this research, Yang was getting skeptical,
but Fermi encouraged his young student, pointing out that, when we are 
students, we have to solve problems, whereas, when we are researchers, we have to ask the right 
sort of questions \cite{Segre}.

It is precisely with this understanding that we have chosen the title for our paper. In order to
help the general reader, we begin by recalling that surfaces have always played an 
important role in general relativity. The mathematical language of general relativity 
was indeed born at the time when Ricci-Curbastro was lecturing on the theory of surfaces at Padova 
University \cite{RicciA,RicciB}. Two- and three-dimensional surfaces have taught us many lessons on
gravitational physics ever since. For example, 
event horizon and apparent horizon of a black hole have, both, 
two-sphere topology \cite{BH1972}, and it is
intriguing that the relation between area $A$ of the event horizon and black hole entropy $S$ (hereafter we set  $G=c=\hbar=k_B=1$):
\begin{equation}
S={A \over 4},
\label{(1.1)}
\end{equation}
can be obtained from the boundary term in the gravitational action of a Schwarzschild black
hole, if the partition function is evaluated at tree level in Euclidean quantum gravity
\cite{GH1977}. The consideration of the variational principle for classical
general relativity \cite{Y1,Y2,CN,OS,HAY} leads again 
to the boundary term used in Ref. \cite{GH1977},
but also to another form of the boundary term, applied by Hartle and Hawking in
quantum cosmology \cite{HH1983}. Furthermore, boundary terms have been considered in 
Ref. \cite{PAD1}, devoted to holography and action functionals, and in Ref. 
\cite{PAD2}, devoted to horizon thermodynamics and emergent gravity. 
Even more recently, the work in Ref. \cite{NB1} has considered the boundary integral 
of $2 \sqrt{-g}(\Theta+\kappa)\sqrt{q}$ for a null boundary, where $\Theta$ is the 
trace of the extrinsic curvature and $\kappa$ is the surface gravity of the null
surface, while $q$ is the determinant of the induced metric on such a surface.
The author of Ref. \cite{NB2} has introduced two new variables to describe
general relativity, applying them to the action principle, and comparing in detail
what results from fixing the induced metric or, instead, the conjugate momentum
at the boundary. The authors of Ref. \cite{NB3} have performed a complete analysis
of the boundary term in the action functional of general relativity when the 
boundary includes null segments in addition to the more usual timelike and 
spacelike segments. Last, but not least, the work in Ref. \cite{NB4} has considered
a spacetime region whose boundary has piecewise $C^{2}$ components, each of which can 
be spacelike, timelike or null, and has obtained a unified treatment of boundary
components by using tetrads.

In a rather different framework, the mathematical community has studied over the centuries the
problem of finding the surface of least area among those bounded by a given curve. In the
attempt of building a rigorous theory of these minimal surfaces, it proved useful to regard a 
hypersurface in ${\bf R}^{n}$ as a boundary of a measurable set $E$, having characteristic 
function 
\begin{equation}
\varphi(x,E)=1 \; {\rm if} \; x \in E, \; 0 \; {\rm if} \; x \in 
{\bf R}^{n}-E
\label{(1.2)}
\end{equation}
whose distributional derivatives have finite total variation. What really matters is then the 
so-called reduced boundary ${\cal F}E$ of $E$. For every $x \in {\cal F}E$ one can
define an approximate normal vector ($A(x,\rho)$ being the open hypersphere of radius $\rho$
centred at $x$)
\begin{equation}
\nu_{\rho}(x)={\int_{A(x,\rho)}{\rm grad} \varphi(x,E) \over 
\int_{A(x,\rho)} \|{{\rm grad} \varphi(x,E)} \| },
\label{(1.3)}
\end{equation}
where $\| v \|$ without subscript denotes in our paper the Euclidean norm of a vector $v$ in 
${\bf R}^{n}$, i.e. $\| v \|=\sqrt{\sum\limits_{i=1}^{n}(v^{i})^{2}}=
\sqrt{\delta_{E}(v,v)}$, $\delta_{E}$ being the Euclidean metric 
${\rm diag}(1,...,1)$. 

It was proved by De Giorgi that if, for some $x \in {\cal F}E$ and some $\rho >0$, the vector
$\nu_{\rho}(x)$ has length close enough to 1, then the difference $1-\| {\nu_{\rho}(x)} \|$ approaches
$0$ as $\rho \rightarrow 0$. This implies in turn that the reduced boundary is analytic in a
neighbourhood of $x$.

In general relativity, measure-theoretic concepts have been exploited 
to obtain very important results, e.g., the proof of the positive-mass 
theorem \cite{SY1,SY2,Bryden} and of the Riemannian Penrose 
inequality \cite{Bray}. Moreover,
in recent years, the concepts and problems of geometric measure theory 
\cite{FF,book1,book2,book3,book4,book5,book6,book7} have been studied in
non-Euclidean spaces and sub-Riemannian manifolds \cite{AP2015}, and the resulting framework
is not only extremely elegant but also conceptually profound.  
Since a physics-oriented reader is not necessarily 
familiar with geometric measure theory, we begin with a pedagogical review of finite-perimeter
sets and reduced boundary properties in Sec. $2$, while their counterpart for Riemannian
manifolds is considered in Sec. $3$. Section $4$ evaluates the general formulae
at the end of Sec. $3$ in Euclidean Schwarzschild geometry and a possible application 
to the study of the intrinsic entropy of a black hole is considered in Sec. $4.1$.
In Sec. $5$, we propose how to exploit the material of Secs. $2$ and $3$ in
order to arrive at a definition of finite-perimeter set and reduced boundary
in paracompact pseudo-Riemannian manifolds.
Concluding remarks and open problems are presented in Sec. $6$, while technical details
are provided in the Appendices.
Our presentation assumes only that the reader is fa\-mi\-li\-ar wi\-th
the ba\-sic elements of Lebesgue's theory of measure and integration 
\cite{Lebesgue,Cafiero,Schwartz,Miranda,APM}. 

\section{Finite perimeter sets in ${\bf R}^{n}$ and their reduced boundary} \label{Sec2}
\setcounter{equation}{0}

Research on geometric measure theory was initiated by Caccioppoli
\cite{RC1,RC2}, who was aware that, in the early fifties, integral
calculus was still lacking a theory of $k$-dimensional integration
in a $n$-dimensional space ($k<n$). He was aiming at a theory, of
the same kind of generality of the Lebesgue theory of $n$-dimensional
integration, relying upon simple and exhaustive notions of 
$k$-dimensional measure and integral, and culminating in an
ultimate extension of integral theorems on differential forms.
Until that time, there had been a variety of efforts, but not
inspired by a clear overall vision, i.e., several definitions of
linear or superficial measure, and various partial extensions of
the Gauss-Green formula. Such a program was carried out and completed
successfully by De Giorgi \cite{DG1,DG2,book2}, 
Federer and Fleming \cite{FF,book1}.
On denoting again by $\varphi(x,E)$ the characteristic function of
a set $E \subset {\bf R}^{n}$ defined in Eq. (1.2), and by $*$
the convolution product of two functions defined on ${\bf R}^{n}$:
\begin{equation}
f*h(x) \equiv \int f(x-\xi) h(\xi) d\xi,
\label{(2.1)}
\end{equation}
De Giorgi defined for all integer $n \geq 2$ and for all
$\lambda >0$ the function
\begin{equation}
\varphi_{\lambda}: x \rightarrow 
\varphi_{\lambda}(x) \equiv (\pi \lambda)^{-{n \over 2}}
{\rm exp} \left(-{\sum\limits_{k=1}^{n}(x_{k})^{2} \over \lambda}\right)
* \varphi(x,E),
\label{(2.2)}
\end{equation}
and, as a next step, the {\it perimeter} of the set
$E \subset {\bf R}^{n}$
\begin{equation}
P(E) \equiv \lim_{\lambda \to 0} \int_{{\bf R}^{n}}
\sqrt{\sum_{k=1}^{n} \left({\partial \varphi_{\lambda}
\over \partial x_{k}}\right)^{2}} \; dx.
\label{(2.3)}
\end{equation}
The perimeter defined in Eq. (2.3) is not always finite. A
necessary and sufficient condition for $P(E)$ to be finite is
the existence of a set function of vector nature completely 
additive and bounded, defined for any set $B \subset {\bf R}^{n}$
and denoted by $a(B)$, verifying the generalized Gauss-Green formula
\begin{equation}
\int_{E}Dh \; dx=-\int_{{\bf R}^{n}}h(x) \; da.
\label{(2.4)}
\end{equation}
If Eq. (2.4) holds, the function $a$ is said to be the distributional
gradient of the characteristic function $\varphi(x,E)$. 
A {\it polygonal domain} is every set $E \subset {\bf R}^{n}$ that
is the closure of an open set and whose topological boundary 
$\partial E$ is contained in the union of a finite number of
hyperplanes of ${\bf R}^{n}$. The sets approximated by polygonal
domains having finite perimeter were introduced by Caccioppoli
\cite{RC1,RC2} and coincide with the collection of all 
finite-perimeter sets \cite{book2}. This is why finite-perimeter 
sets are said to be Caccioppoli sets. 

The modern presentation of these concepts is even more refined.
If $u$ is a Le\-bes\-gue-sum\-ma\-ble func\-ti\-on on an open set $\Omega$ of ${\bf R}^{n}$, 
$u$ is said to have {\it bounded variation} in $\Omega$ if its distributional
derivative is representable by a measure in $\Omega$, in such a way that 
(cf. Eq. (2.4)) one can write \cite{book5}
\begin{equation}
\int_{\Omega}u {\partial \phi \over \partial x^{i}}
=-\int_{\Omega} \phi d(D_{i}u), \; \; \forall \phi \in C_{c}^{\infty}(\Omega),
\label{(2.5)}
\end{equation}
where $Du=(D_{1}u,...,D_{n}u)$ is a ${\bf R}^{n}$-valued measure. The {\it variation}
of $u$ in $\Omega$ is a measure, denoted by $|Du|$, which, when evaluated on $\Omega$, gives:
\begin{equation}
|Du|(\Omega) \; \equiv \; {\rm sup} \left \{
\int_{\Omega}u \; {\rm div}h \; dx \;  : 
h \in [C_{c}^{1}(\Omega,{\bf R}^{n})], \; \left \| h \right \|_{L^{\infty}} 
\leq 1 \right \},
\label{(2.6)}
\end{equation}
where $\left \| h \right \|_{L^{\infty}}$ 
is the essential supremum norm on all components of $h$, i.e., 
$\left \| h \right \|_{L^{\infty}}
= \inf \left\{ c \geq 0 ~|~ |h_i(x)| \leq c ~ {\rm for \; almost \; every} 
\; x \in \Omega, \forall i = 1,\dots,n\right\}$.
Note that integration by parts yields $\int_\Omega u {\rm div} h = - \int_\Omega ({\rm grad} u) h$. 
Hence, calculating the sup and applying Cauchy--Schwartz 
one obtains the simple but important relation
$ |Du|(\Omega) = \left \| {\rm grad}u \right \|_{L^{1}(\Omega)}$.

The {\it perimeter} of a Lebesgue measurable set $E$ in $\Omega$ is the variation of the
characteristic function $\varphi_{E}$ (hereafter we denote $\varphi(x,E)$ by
$\varphi_{E}$ for simplicity of notation), i.e. (cf. Eq. (2.6))
\begin{equation}
P(E,\Omega) \equiv |D \varphi_{E} | (\Omega)
={\rm sup} \left \{ \int_{\Omega} {\rm div}h \; dx \;  :
h \in [C_{c}^{1}(\Omega,{\bf R}^{n})], \; \left \| h \right \|_{L^{\infty}}
\leq 1 \right \},
\label{(2.7)}
\end{equation}
while the associated reduced boundary, denoted by ${\cal F}E$, is the set
of points of the topological boundary $\partial E$ satisfying the condition
\begin{equation}
{\rm there} \; {\rm exists} \; \nu_{E}(x) \equiv
\lim_{\rho \to 0^{+}} {D \varphi_{E} (A(x,\rho)) \over
|D \varphi_{E} | (A(x,\rho))},
\label{(2.8)}
\end{equation}
jointly with the unit Euclidean norm condition 
\begin{equation}
\left \| \nu_{E} \right \| = 1.
\label{(2.9)}
\end{equation}
The unit norm of the generalized
inner normal $\nu_{E}$ makes it possible to obtain the identity
\begin{eqnarray}
\; & \; & 
{1 \over 2 |D\varphi_{E}|(A(x,\rho))}
\int_{A(x,\rho)}  \| {\nu_{E}(y)-\nu_{E}(x)} \|^{2}
d |D \varphi_{E}|(y)
\nonumber \\ 
&=& 1- \left \langle \nu_{E}(x),{D \varphi_{E}(A(x,\rho)) \over
|D \varphi_{E}| (A(x,\rho))} \right \rangle .
\label{(2.10)}
\end{eqnarray}
By inserting the definition~(\ref{(2.8)}), and using the unit norm 
condition~(\ref{(2.9)}), one finds that 
\begin{equation}
\left \langle \nu_{E}(x),{D \varphi_{E}(A(x,\rho)) \over
|D \varphi_{E}| (A(x,\rho))} \right \rangle 
=
\lim_{\rho' \to 0^{+}} \left \langle {D \varphi_{E} (A(x,\rho')) \over
|D \varphi_{E} | (A(x,\rho'))} ,{D \varphi_{E}(A(x,\rho)) \over
|D \varphi_{E}| (A(x,\rho))} \right \rangle \rightarrow 1 ,
\label{(2.11)}
\end{equation}
as $\rho$ approaches $0$,
and therefore an equivalent definition of the inner normal is
\begin{eqnarray}
\; & \; &
\lim_{\rho \to 0} {1 \over  |D\varphi_{E}|(A(x,\rho))}
\int_{A(x,\rho)} \| {\nu_{E}(y)-\nu_{E}(x)} \|^{2}
d |D \varphi_{E}|(y) =0,
\nonumber \\
& \; & \left \| \nu_{E}(x) \right \|  = 1 .
\label{(2.12)}
\end{eqnarray}
The above relation will be crucial in the next section, 
because it will tell us how to define 
the concept of reduced boundary in Riemannian geometry. 
The theory of finite-perimeter sets was initiated in the works by Caccioppoli
\cite{RC1,RC2}, but it was De Giorgi who put on firm ground the brilliant ideas of
Caccioppoli in an impressive series of theorems \cite{DG1,DG2}. 

In order to help the general reader, we give below a standard but very useful
example of reduced boundary. For this purpose, given the finite-perimeter set
\begin{equation}
E \equiv \left \{ (x,y): 0 \leq x,y \leq 1 \right \} \cup 
\left \{ (x,0): -1 \leq x \leq 1 \right \} \subset {\bf R}^{2},
\label{(2.13)}
\end{equation}
which represents (see Fig. 1) a square jointly with a line segment sticking out
on the left, one finds that its perimeter $P(E)=4$, which ignores the additional
line segment. However, the topological boundary
\begin{eqnarray}
\partial E &=& \left \{(x,0): -1 \leq x \leq 1 \right \}
\cup \left \{ (x,1): 0 \leq x \leq 1 \right \}
\nonumber \\
& \cup & \left \{ (x,y): x \in \{0,1 \}, 0 \leq y \leq 1 \right \}
\label{(2.14)}
\end{eqnarray}
has one-dimensional Hausdorff measure equal to $5$. Thus, the appropriate
boundary should be a subset of the topological boundary, because the Hausdorff 
measure \cite{book6} of the latter overcompensates for the perimeter $P(E)$.

\begin{figure}
\centering
\includegraphics[height=6cm]{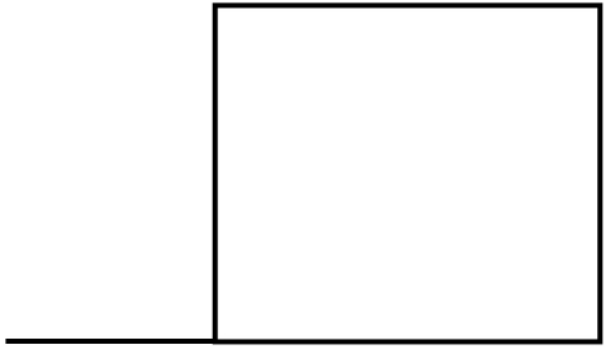}
\caption{The set $E$ defined in Eq. (2.13).}
\end{figure}

\section{Reduced boundary in Riemannian geometry} \label{Sec3}
\setcounter{equation}{0}

The passage to Riemannian geometry makes it necessary to take into account some
additional concepts, for which we rely mainly on the work in Ref. \cite{AP2015}. 
Thus, we consider a smooth, oriented, connected, $n$-dimensional manifold $M$ with
tangent bundle $TM$, endowed with a smooth $n$-form $\omega$ and associated
volume measure
\begin{equation}
m(E)=\int_{E} \omega ,  \; \; E \subset M.
\label{(3.1)}
\end{equation}
If $X: M \rightarrow TM$ is a smooth vector field and $f$ is a function of class 
$C^{1}$, one has
\begin{equation}
D_{X}f = (Xf)m.
\label{(3.2)}
\end{equation}
Since we are dealing with Riemannian geometry, we can exploit
the existence of a Riemannian metric $g$ to define the spaces
\begin{equation}
{\cal D}(x) \equiv \left \{ v \in T_{x}M \; \; \;  :  g(v,v)  < \infty \right \},
\label{(3.3)}
\end{equation}
i.e., the family of all vectors belonging to the tangent space to $M$ at $x$ 
and having finite norm.
If $\Omega$ is an open set in $M$, the space of smooth sections of ${\cal D}$ is
denoted by $\Gamma(\Omega,{\cal D})$ and is defined by
\begin{equation}
\Gamma(\Omega,{\cal D}) \equiv \left \{ Y \;  :  Y \; {\rm is} \;
{\rm smooth} \; {\rm in} \; \Omega, \; Y(x) \in {\cal D}(x), \;
\forall x \in \Omega \right \}, 
\label{(3.4)}
\end{equation}
where smoothness can be taken to be of class $C^{k}$, up to $k=\infty$.
We shall need a subset of the space of smooth sections of ${\cal D}$, defined as
\begin{equation}
\Gamma^{g}(\Omega,{\cal D}) \equiv \left \{ Y \in \Gamma(\Omega,{\cal D}) \;  :
g(Y(x),Y(x)) \leq 1, \; \forall x \in \Omega \right \}.
\label{(3.5)}
\end{equation}
At this stage, we can say that if the set $\Omega \subset M$ is open and the function
$u$ is Lebesgue summable on $(\Omega,m)$, such an $u$ has {\it bounded variation}
in $\Omega$ if $D_{X}u$ (see (3.2)) exists for all vector fields 
$X \in \Gamma^{g}(\Omega,{\cal D})$, and
\begin{equation}
{\rm sup} \; \left \{ |D_{X}u |(\Omega) \;  :  X \in
\Gamma^{g}(\Omega,{\cal D}) \right \} < \infty,
\label{(3.6)}
\end{equation}
where the variation of $D_{X}u$ in $\Omega$ is
\begin{equation}
|D_{X}u|(\Omega) \equiv {\rm sup} \left \{ \int_{\Omega} 
u \; {\rm div}_{\omega}(f X)\omega \;  :  f \in
C_{c}^{\infty}(\Omega), \; \left \| f \right \|_{L^{\infty}} \leq 1 \right \},
\label{(3.7)}
\end{equation}
where now $\Supnorm{f} \leq 1$, for a scalar function $f$, 
just means that $|f(x)|\leq 1$ almost everywhere.
When we write ${\rm div}_{\omega}(f X)\omega $ we mean 
the divergence of the vector field $f X$, which
in Riemannian geometry can be defined as the  Lie derivative 
with respect to $X$ of the volume form: $\pounds_X (\omega) 
= ( {\rm div}_\omega X) \, \omega $.

If the condition (3.6) is fulfilled, one writes 
that the function $u$ belongs to the functional space
$$
BV(\Omega,g,\omega).
$$
Once that an orthonormal frame $(X_{1},...,X_{m})$ is fixed, one can define a measure
$|D_{g} u|$ which is the total variation of the vector measure \cite{AP2015}
\begin{equation}
X u \equiv \left(D_{X_{1}}u,...,D_{X_{m}}u \right ).
\label{(3.8)}
\end{equation}
In other words, one has the local representation (see (3.7)) of the measure in the form
\begin{equation}
|D_{g}u|  = {\rm sup} \left \{ |D_{X}u | (\Omega) \;  :  X \in \Gamma^{g}(\Omega,{\cal D}),
\; \left \| X \right \|_{L^{\infty}} \leq 1 \right \}.
\label{(3.9)}
\end{equation}
where the metric $g$ in $D_g u$ occurs in the definition because $X \in \Gamma^g(\Omega, \mathcal D)$.

The action of the vector field $X$ on the characteristic function $\varphi_{E}$ is
expressed by the decomposition
\begin{equation}
X \varphi_{E}=\nu_{E}^{*} \; |D_{g}\varphi_{E}| ,
\label{(3.10)}
\end{equation}
where $|D_{g}\varphi_{E}|$ is defined by Eq. (3.9) with $u$ replaced
by the characteristic function, and the dual normal to $E$
$$
\nu_{E}^{*}: \Omega \rightarrow {\bf R}^{n}
$$
is a Borel vector field (see Appendix A) with unit norm. 
Such a decomposition is said to be the {\it polar decomposition}.

The {\it reduced boundary} of $E$, denoted by ${\cal F}_{g}^{*}E$, is the set of all points
in the support of $|D_{g} \varphi_{E}|$ satisfying the 
conditions \cite{AP2015}
\begin{equation}
\lim_{r \to 0^{+}} {\rm inf} \;
{{\rm min} \left \{ m(A(x,r) \cap E),m(A(x,r) \setminus E) \right \}
\over m(A(x,r))} >0,
\label{(3.11)}
\end{equation}
\begin{equation}
\lim_{r \to 0^{+}} {\rm sup} 
{|D_{g} \varphi_{E}| (A(x,r)) \over
{1 \over r} m(A(x,r))} < \infty,
\label{(3.12)}
\end{equation}
and (cf. Eq. (2.12))
\begin{equation}
\lim_{r \to 0^{+}} 
{1 \over |D_{g} \varphi_{E}|(A(x,r))}
\int_{A(x,r)}
\| \nu_{E}^{*}(y) - \nu_{E}^{*}(x) \|^{2}
\; d |D_{g} \varphi_{E}| (y)=0,
\label{(3.13)}
\end{equation}
the squared norm in the integrand being now the squared Riemannian norm.
Such a concept of reduced boundary is independent of the choice of orthonormal frame, unlike 
the dual normal $\nu_{E}^{*}$ \cite{AP2015}.
Equations (3.11)-(3.13) can be considered because the rectifiability properties of the
reduced boundary are local, and hence can be formulated in terms of geodesic balls
of small radius.

It should be stressed that the dual normal depends non-smoothly 
on $x$ as $x$ moves on the support of the measure 
$|D_{g}\varphi_{E}|$. This is why Eq. (3.13) expresses a non-trivial
property and deserves a careful check, that we perform in the
next section. For further material on the concept of reduced boundary,
we refer the reader to Appendix B.

\section{Application to Euclidean Schwarzschild geometry}
\setcounter{equation}{0}

The example at the end of Sec. \ref{Sec2} shows that, for a given finite-perimeter
set, the concept of reduced boundary may be more relevant. Moreover,
as far as we know, the general formulae at the end of Sec. \ref{Sec3} have been never evaluated
in gravitational physics. Thus, we here consider a first example, provided by the
so-called Euclidean Schwarzschild geometry \cite{GW1980}, where the $4$-metric 
\begin{equation}
g={\rm diag} \left( \left(1-{2M \over r} \right),
\left(1-{2M \over r}\right)^{-1},r^{2},r^{2}\sin^{2}\theta \right)
\label{(4.1)}
\end{equation}
is positive-definite and the $g_{00}$ component has therefore 
opposite sign with respect to the Lorentzian
signature case of general relativity. 
With reference to Eqs. (\ref{(3.11)})-(\ref{(3.13)}), our set $A$ is the open
ball of radius $s$ centred at $x$. The integrand in Eq. (3.13) involves
different points $x$ and $y$, at which we use $x_{0},r,\theta,\phi$ coordinates, 
and we exploit spherical symmetry and static nature of $g$ to assume
that the metric \eqref{(4.1)} takes the same value at $x$ and at $y$.
Hence the integrand in (3.13) reads as
(cf. comments following Eq. (1.3))
\begin{eqnarray}
\; & \; & \| \nu^{*}(y)-\nu^{*}(x) \|^{2}
=g(\nu^{*}(y)-\nu^{*}(x),\nu^{*}(y)-\nu^{*}(x))
\nonumber \\
&=& \sum_{\lambda,\mu=0}^{3}
g_{\lambda \mu} \Bigr({\nu^{*}}^{\lambda}(y)-{\nu^{*}}^{\lambda}(x)\Bigr)
\Bigr({\nu^{*}}^{\mu}(y)-{\nu^{*}}^{\mu}(x)\Bigr)
\nonumber \\
&=& g(\nu^{*}(y),\nu^{*}(y))+g(\nu^{*}(x),\nu^{*}(x))
\nonumber \\
&-& \sum_{\lambda,\mu=0}^{3} g_{\lambda \mu}
\Bigr({\nu^{*}}^{\lambda}(y){\nu^{*}}^{\mu}(x)
+{\nu^{*}}^{\lambda}(x){\nu^{*}}^{\mu}(y)\Bigr)
\nonumber \\
&=& 2-2 g(\nu^{*}(x),\nu^{*}(y)),
\label{(4.2)}
\end{eqnarray}
where we have exploited the symmetry of $g$ and the fact that the dual normal
$\nu^{*}$ must have unit norm \cite{AP2015}: 
\begin{equation}
g(\nu^{*},\nu^{*})=\sum_{\lambda,\mu=0}^{3}g_{\lambda \mu}
{\nu^{*}}^{\lambda}(x) {\nu^{*}}^{\mu}(x)
=\sum_{\lambda,\mu=0}^{3}g_{\lambda \mu}
{\nu^{*}}^{\lambda}(y) {\nu^{*}}^{\mu}(y)=1.
\label{(4.3)}
\end{equation}
Such a condition is fulfilled by
\begin{equation}
{\nu^{*}}^{\lambda}(x)={\psi^{\lambda}(x) \over \sqrt{g_{\lambda \lambda}}}
\; \; \forall \lambda=0,1,2,3,
\label{(4.4)}
\end{equation}
where the $\psi^{\lambda}$ are Borel functions (see Appendix A), and
hence takes eventually the form
\begin{equation}
\sum_{\lambda=0}^{3}(\psi^{\lambda}(x))^{2}=1.
\label{(4.5)}
\end{equation}
The explicit evaluation of the limit in Eq. (3.13) yields therefore
\begin{eqnarray}
l & \equiv & \lim_{s \to 0^{+}} 
{\int_{A(x,s)} \| \nu^{*}(y)-\nu^{*}(x) \|^{2}
d |D_{g}\varphi_{E}|(y)\over |D_{g}\varphi_{E}|(A(x,s))}
\nonumber \\
&=& 2-2 \lim_{s \to 0^{+}}
{\int_{A(x,s)}g(\nu^{*}(x),\nu^{*}(y)) 
d |D_{g}\varphi_{E}|(y) \over
|D_{g}\varphi_{E}|(A(x,s))},
\label{(4.6)}
\end{eqnarray}
where
\begin{equation}
g(\nu^{*}(x),\nu^{*}(y))=\sum_{\lambda=0}^{3}
\psi^{\lambda}(x) \psi^{\lambda}(y).
\label{(4.7)}
\end{equation}
This leads in turn to the formula
\begin{equation}
l=2-2 \lim_{s \to 0^{+}} 
\sum_{\lambda=0}^{3}\psi^{\lambda}(x)
{\int_{A(x,s)}\psi^{\lambda}(y)d |D_{g} \varphi_{E}|(y) \over
|D_{g} \varphi_{E}| (A(x,s))},
\label{(4.8)}
\end{equation}
and hence the limit $l$ vanishes since it takes eventually the same 
functional form as in flat Euclidean $4$-space (cf. 
Sec. $3.2$ of Ref. \cite{book4}, and our Eqs. (2.10) and (2.11)).

In conclusion, since the general formulae at the end of Sec. \ref{Sec3} 
have been here shown to be computable in a case of interest for 
gravitational physics and Euclidean quantum gravity, this technical result 
looks rather helpful for gaining familiarity with reduced-boundary calculations. 

\subsection{An implication for the theory of quantum gravity: the entropy of a black hole}

It is well-known  that the most general form of the action $\mathcal{I}$  of the gravitational field over a region $\mathcal{B}$ of the spacetime  having topological boundary $\partial \mathcal{B}$ includes, apart from the usual Einstein-Hilbert term, the Gibbons-Hawking-York boundary term \cite{GH1977,Y1,Y2}, which encompasses  the sum of the trace $K$ of the extrinsic curvature of $\partial \mathcal{B}$ and a term $C$ depending only on the induced metric $h_{\mu \nu}$ on $\partial \mathcal{B}$. This means that, for generic geometries, $\mathcal{I}$ assumes the form (following the same conventions as in Ref. \cite{GH1977})
\begin{equation} \label{generic action_1}
\mathcal{I} = \dfrac{1}{16 \pi }\int_{\mathcal{B}} \sqrt{-g}\, d^4x \,R \;+\; \int_{\partial \mathcal{B}} \sqrt{-h}\, d^3x \, D , 
\end{equation}
with
\begin{equation} \label{generic action_2}
D \equiv \dfrac{1}{8 \pi}K(g) +C(h),
\end{equation}
where we have stressed the dependence of $K$ on the metric $g_{\mu \nu}$ and of $C$ on the induced metric $h_{\mu \nu}$.

As pointed out in Sec. 1, whenever the boundary $\partial \mathcal{B}$ is  non-null, the Gibbons-Hawking-York boundary term is essential for the path-integral approach to the quantization of the gravitational field \cite{Hawking} and plays an important role in the evaluation of the intrinsic quantum mechanical entropy of a black hole (see Eq. (\ref{(1.1)})) \cite{GH1977}. In the case of asymptotically flat spacetimes, e.g. Schwarzschild,  the Gibbons-Hawking-York boundary term involves the difference between the  trace $K$ of the extrinsic curvature of $\partial \mathcal{B}$ in the metric $g_{\mu \nu}$ and the trace $K_0$ of the extrinsic curvature of the boundary $\partial \mathcal{B}$ imbedded in flat spacetime. This means that, for asymptotically flat spacetimes, Eq. (\ref{generic action_2}) assumes a different form and hence the action $\mathcal{I}$ can be written as 
\begin{equation} \label{action}
\mathcal{I} = \dfrac{1}{16 \pi }\int_{\mathcal{B}} \sqrt{-g}\, d^4x \,R \;+\; \dfrac{1}{8 \pi } \left[\int_{\partial \mathcal{B}} \sqrt{-h}\, d^3x \,K \; - \; \int_{\partial \mathcal{B}} \sqrt{-h}\, d^3x \,K_0\right]. 
\end{equation}

At this stage, let us evaluate (\ref{action}) within the framework spelled out in Sec. 3. 
This unavoidably leads us to  deal with a new research field whose rich structure  has  never been exploited, as far as we know, in  gravitational  settings. In particular, in this context, the usual definition of normal vector field to a surface is superseded by the novel concept of dual normal, which we indicate with  $\nu^{* \mu}$ (see Eqs. (\ref{(4.4)})  and (\ref{(4.5)})). Such  normal does not display the usual  interpretation considered in differential geometry and, among the other features, it can also be a discontinuous function. Therefore, we propose an original method to calculate, within this new formalism having deep and unexplored potentialities, the entropy of a black hole. For this reason, in Eq. (\ref{action}) we will replace the region  $\mathcal{B}$ and its \emph{topological} boundary $\partial \mathcal{B}$  with a finite-perimeter region and its \emph{reduced} boundary, respectively. If we suppose that the components of $\nu^{* \mu}$,  given in Eqs. (\ref{(4.4)}) and (\ref{(4.5)}), are such that the  Borel functions $\psi^\mu(x)$  are all nonvanishing and  depend only on the radial coordinate $r$, we find that, for the Euclidean Schwarzschild geometry (\ref{(4.1)}), 
\begin{eqnarray}
\nabla_\mu \nu^{* \mu} &=& \dfrac{1}{\sqrt{g}} \partial_\mu \left(\sqrt{g} \,\nu^{*\mu}\right)
\nonumber \\
&=& \dfrac{\sqrt{g_{rr}}}{g_{\theta \theta}} \left[\dfrac{g_{\theta \theta}}{g_{rr}} \partial_r \psi^r + M \psi^r \right] + \dfrac{2}{r}\dfrac{\psi^r}{\sqrt{g_{rr}}} + \cot \theta \dfrac{\psi^\theta}{\sqrt{g_{\theta \theta}}},
\end{eqnarray}
where we recall that (cf. Eq. (\ref{(4.1)}))
\begin{eqnarray} \label{grr-gthetatheta}
 g_{rr} & = & \left(1-\dfrac{2M}{r}\right)^{-1}, \\
 g_{\theta \theta} &=& r^2.
\end{eqnarray}
Bearing in mind the above equations, we find that
\begin{eqnarray} \label{integral-of-nabla-nu}
\int_{r=r_C}  && \sqrt{-h}\, d^3x \, \nabla_\mu \nu^{* \mu} 
\nonumber \\ 
&=& - i  \left[ \dfrac{g_{\theta \theta}}{g_{rr}} \partial_r \psi^r +\psi^r \left(M+\dfrac{2}{r}\dfrac{g_{\theta \theta}}{g_{rr}}\right) \right]_{r=r_C} \int_0^{8 \pi M} d x_0 \int_0^\pi d\theta \sin \theta \int_0^{2 \pi} d \phi
\nonumber \\ 
&-&i \left[\sqrt{\dfrac{g_{\theta \theta}}{g_{rr}}} \, \psi^\theta \right]_{r=r_C} \int_0^{8 \pi M} d x_0 \int_0^\pi d\theta \cos \theta \int_0^{2 \pi} d \phi
\nonumber \\ 
&=& -32  \pi^2 i M \left[ \dfrac{g_{\theta \theta}}{g_{rr}} \partial_r \psi^r + \psi^r \left(M + \dfrac{2}{r}\dfrac{g_{\theta \theta}}{g_{rr}} \right) \right]_{r=r_C},
\end{eqnarray}
 where we have taken into account that the Euclidean Schwarzschild metric entails a periodic Euclidean time $x_0$  with period $8 \pi M$ (i.e., the inverse Hawking temperature). The integral (\ref{integral-of-nabla-nu}) has been evaluated on the hypersurface $r=r_C$ ($r_C$ being a constant) having topology $S^1 \times S^2$ \cite{GH1977}.  Such hypersurface is assumed to define the boundary of a finite-perimeter region of the Euclidean Schwarzschild spacetime. Since for this geometry the scalar curvature vanishes, from Eq. (\ref{action}) we only need to consider the term 
\begin{eqnarray} \label{final_expression}
\tilde{\mathcal{I}} &\equiv & \dfrac{1}{8 \pi } \left[\int_{r=r_C} \sqrt{-h}\, d^3x \,\nabla_\mu \nu^{* \mu} \; - \; \int_{r=r_C} \sqrt{-h}\, d^3x \,\nabla_\mu \nu^{* \mu}_0 \right] 
 \nonumber \\
&=& -4 \pi i M \Biggl[ \dfrac{g_{\theta \theta}}{g_{rr}} \partial_r \psi^r + \psi^r \left(M + \dfrac{2}{r}\dfrac{g_{\theta \theta}}{g_{rr}} \right)
\nonumber \\
 &+&   \left(\dfrac{-g_{\theta \theta}}{\sqrt{g_{rr}}}\right) \left(\partial_r \psi^r_0 + \dfrac{2}{r} \psi^r_0 \right)\Biggr]_{r=r_C}, 
\end{eqnarray}
where we have taken into account that 
\begin{equation}
\nabla_\mu \nu^{* \mu}_0 =\partial_r \psi^r_0 + \dfrac{2}{r} \psi^r_0  +\dfrac{\cot \theta}{r} \psi^\theta_0. 
\end{equation}

Note that our Eq. (\ref{final_expression}) should be compared with the standard expression (\ref{action}), which  
in Euclidean Schwarzschild geometry reads as
\begin{eqnarray} \label{standard-expression}
\mathcal{I} &=&  \dfrac{1}{8 \pi } \left[\int_{\partial \mathcal{B}} \sqrt{-h}\, d^3x \,K \; - \; \int_{\partial \mathcal{B}} \sqrt{-h}\, d^3x \,K_0\right] 
\nonumber \\
&=& -4 \pi i M \left[ (2r-3M) - {2r}\sqrt{1-\dfrac{2M}{r}} \right]_{r=r_C}. 
\end{eqnarray}
Thus, Eqs. (\ref{grr-gthetatheta}), (\ref{final_expression}) and (\ref{standard-expression}) imply that
\begin{eqnarray} \label{I_tilde-I}
\dfrac{i\left(\tilde{\mathcal{I}}-\mathcal{I}\right)}{4 \pi M}  &=&  \Biggl[r^2\sqrt{1-\dfrac{2M}{r}} \left(\sqrt{1-\dfrac{2M}{r}} \partial_r \psi^r -\partial_r \psi^r_0\right) 
\nonumber \\
&+&\left(2r-3M\right) \left(\psi^r-1\right)-2r \sqrt{1-\dfrac{2M}{r}} \left(\psi^r_0-1 \right) \Biggr]_{r=r_C}.
\end{eqnarray}
In the above equation, the standard case treated in Ref. \cite{GH1977} is immediately recovered if $\psi^r=\psi^r_0=1$, since in this case Eq. (\ref{I_tilde-I}) gives readily $\tilde{\mathcal{I}}=\mathcal{I}$. In this sense, the action $\tilde{\mathcal{I}}$, defined in Eq. (\ref{final_expression}), can be interpreted as a generalization of the Euclidean action $\mathcal{I}$, given in Eq. (\ref{standard-expression}). 

Starting from Eq. (\ref{final_expression}), by employing standard field-theoretic techniques along with general thermodynamic arguments \cite{GH1977} and by means of the generalized Smarr formula \cite{Smarr1973}, it is possible to evaluate the entropy $\tilde{S}$ of a black hole. In this way, we easily find
\begin{equation} \label{corrected-entropy}
\tilde{S}= \dfrac{A}{2} \left[1+\dfrac{\mathscr{F}\left(\psi\right)}{2M}\right],
\end{equation}
where (cf. Eqs. (\ref{grr-gthetatheta}) and (\ref{final_expression}))
\begin{eqnarray}
\mathscr{F}\left(\psi\right) &\equiv & \Biggl[ r^2 \left(1-\dfrac{2M}{r}\right) \partial_r \psi^r + \left(2r-3M\right) \psi^r 
\nonumber \\
&-& r \sqrt{1-\dfrac{2M}{r}} \left(r \, \partial_r \psi^r_0 -2 \psi^r_0 \right)\Biggr]_{r=r_C}.
\end{eqnarray}
The above equations enable one to infer  the corrections to the standard formula (\ref{(1.1)}), which can be easily obtained, since whenever $\psi^r=\psi^r_0=1$, we have
\begin{equation} \label{limit-of-F-psi}
\mathscr{F}\left(\psi\right)^{\rm (standard)} = \left(2r-3M\right) -2r\sqrt{1-\dfrac{2M}{r}} = -M + {\rm O}(M^2 r_C^{-1}),
\end{equation}
and hence Eq. (\ref{corrected-entropy}) reduces to Eq. (\ref{(1.1)})  if higher-order corrections in (\ref{limit-of-F-psi}) are neglected. 

Some comments on the results displayed in this section are in order. It is clear that Eq. (\ref{corrected-entropy}) gives rise to a general expression for the entropy $\tilde{S}$ and 
that a specific calculation   requires the knowledge of the Borel functions $\psi^r$ and $\psi^r_0$ occurring  in $\nu^{*\mu}$ and $\nu^{*\mu}_0$, respectively.  So far, we have implicitly assumed that the dual normal is at least piecewise differentiable. This means that  the integrals analyzed in this section  should be evaluated on a family of sets whose union gives the hypersurface $r=r_C$. Therefore, within our framework both the entropy and the temperature of the black hole can exhibit discontinuities. In the case in which the components $\psi^\mu(x)$ are not piecewise differentiable, the derivatives of dual normal $\nu^{* \mu}$ cannot even be defined. 
In particular,  these functions might not admit  a Taylor expansion, unlike the standard case treated in Ref. \cite{GH1977} (see Eq. (\ref{limit-of-F-psi})). 

These interesting topics deserve consideration in a separate paper.

\section{Riemannian metrics in pseudo-Riemannian geometry}
\setcounter{equation}{0}

The material presented so far is a measure-theoretic formulation of the concepts of normal vector, 
finite-perimeter sets and their boundary, and at first sight it might seem that no obvious 
counterpart can be conceived in pseudo-Riemannian geometry.
However, we may recall a theorem \cite{GH1979} according to which a 
manifold $M$ admits a positive-definite
metric $\gamma$ if and only if it is paracompact. The proof consists of first choosing, for
each of a countable collection of coordinate patches that cover $M$, 
a metric that is positive-definite in 
the interior of the patch and zero outside, and then taking the sum of these metrics, possibly 
after rescaling so that the sum converges to some metric on $M$. Moreover, if $M$ is endowed
with a direction vector field $\xi$, the metric with components
\begin{equation}
g_{\alpha \beta}=\gamma_{\alpha \beta}-2 \sum_{\rho,\sigma=0}^{3}
{(\gamma_{\alpha \rho}\xi^{\rho}) (\gamma_{\beta \sigma} \xi^{\sigma})
\over \gamma(\xi,\xi)}
\label{(5.1)}
\end{equation}
is Lorentzian and is independent of the scaling of $\xi$. 

On a manifold there exists indeed an uncountable infinity of Riemannian
metrics, and infinitely many partitions of unity 
$\left \{ \rho_{\alpha} \right \}$ that can be used to glue such metrics
and obtain a global metric \cite{Abate}. 
Of course, this would be of little help without a strategy
for choosing a definite Riemannian metric $\gamma$. For this purpose,
we consider again Eq. (5.1) and point out that, once a Lorentzian metric 
is given, we have to find the components of the timelike vector field
that are consistent with the choice of $\gamma_{\alpha \beta}$ and
$g_{\alpha \beta}$ in the equation
\begin{equation}
\gamma_{\alpha \beta}-g_{\alpha \beta}=2 \sum_{\rho,\sigma=0}^{3}
{(\gamma_{\alpha \rho}\xi^{\rho})(\gamma_{\beta \sigma}\xi^{\sigma})
\over \gamma(\xi,\xi)}.
\label{(5.2)}
\end{equation}
Two cases of diagonal metrics $\gamma$ and $g$ are here brought to the
attention of the reader.
\vskip 0.3cm
\noindent
(i) {\it Euclidean metric in Minkowski space-time}. If we choose
$$
\gamma_{\alpha \beta}={\rm diag}(1,1,1,1), \; \;
g_{\alpha \beta}={\rm diag}(-1,1,1,1),
$$
Eq. (5.2) yields
\begin{equation}
2=2 {(\xi^{0})^{2}\over \gamma(\xi,\xi)},
\label{(5.3)}
\end{equation}
\begin{equation}
0=2{(\xi^{k})^{2}\over \gamma(\xi,\xi)} \Longrightarrow 
\xi^{k}=0, \; \forall k=1,2,3,
\label{(5.4)}
\end{equation}
jointly with equations for $\gamma_{0k}-g_{0k}$, 
$\gamma_{ij}-g_{ij}$ that are identically satisfied by virtue of
Eq. (5.4). Hence we find
\begin{equation}
\xi=\xi^{0}{\partial \over \partial x^{0}}
\; \; \; \; 
\forall \xi^{0} \in {\bf R}- \left \{ 0 \right \},
\label{(5.5)}
\end{equation}
which is of course timelike in the Minkowski metric.
\vskip 0.3cm
\noindent
(ii) {\it Euclidean Schwarzschild metric in Lorentzian-signature
Schwarzschild space-time}. If we assume that $\gamma_{\alpha \beta}$
is the Riemannian metric with the components in Eq. \eqref{(4.1)}, while
$$
g_{\alpha \beta}={\rm diag} \left(-\left(1-2{M \over r}\right),
\left(1-2{M \over r}\right)^{-1},r^{2},r^{2}\sin^{2}\theta \right),
$$
we find, from Eq. (5.5), the system of equations
\begin{equation}
2 \left(1-2{M \over r}\right)=2 
{(\gamma_{00}\xi^{0})^{2}\over \gamma(\xi,\xi)},
\label{(5.6)}
\end{equation}
and (with no summation over $k$)
\begin{equation}
0=2 {(\gamma_{kk}\xi^{k})^{2}\over \gamma(\xi,\xi)}
\Longrightarrow \xi^{k}=0, \; \forall k=1,2,3,
\label{(5.7)}
\end{equation}
completed by equations for $\gamma_{0i}-g_{0i}$, 
$\gamma_{ij}-g_{ij}$, that are identically satisfied by
virtue of Eq. (5.7). Hence we find again the desired vector
field in the form (5.5).

\section{Concluding remarks and open problems}
\setcounter{equation}{0}

Geometric measure theory can be viewed as a version of differential geometry
studied with the help of measure theory, in order to deal with maps and
surfaces that are not necessarily smooth, with applications to the calculus 
of variations \cite{book6,Morgan}. In our paper  
we have focused on the task of evaluating the
integral formula (\ref{(3.13)}) for a finite-perimeter set that is a portion
of Riemannian Schwarzschild geometry.

As far as we can see, in the course of considering a measure-theoretic approach to the
boundary concept in pseudo-Riemannian geometry, one has to take into account
the infinitely many Riemannian metrics $\gamma$ that are available, as we have 
shown in Sec. $5$.
If we agree that the metric $\gamma_{B}$ is equivalent to the
metric $\gamma_{A}$ if a diffeomorphism turns $\gamma_{A}$ into $\gamma_{B}$,
the resulting reduced boundary is actually an equivalence class.
More precisely, to each representative of the same equivalence class of Riemannian 
metrics, we associate the same set of points that form the reduced boundary of a
finite-perimeter set $S$. Two Riemannian metrics not related by a diffeomorphism 
belong instead to different equivalence classes of Riemannian metrics, and the
associated reduced boundaries of $S$ are inequivalent.

The method suggested by our Eqs. (5.2)-(5.7) consists of associating a
Riemannian metric to the original Lorentzian metric, 
{\it whenever the physical space-time is endowed with a globally defined
timelike vector field}, and then exploiting
our work in Sec. $4$ in the Schwarzschild case, and the huge amount of
rigorous results of Refs. \cite{book1,book2,book3,book4,book5,book6} in the
Euclidean case. This framework might have important implications for a
functional integral approach to quantum gravity, since one might try
to obtain the in-out amplitude \cite{Hawking} 
by integrating only over Riemannian manifolds
with finite perimeter that are obtained from a Lorentzian counterpart 
in the sense of section $5$ and have a non-empty
reduced boundary, for which therefore the condition (\ref{(3.13)}) is fulfilled.
In other words, finite-perimeter sets are a good candidate for the
functional integration because they are reasonably well understood, and their
reduced boundary is what really matters, unlike the topological boundary
(see again the simple example discussed at the end of Section $2$).
The action functional to be used in Euclidean quantum theory might be therefore 
taken over a finite-perimeter portion of Riemannian 4-geometry, and 
supplemented by a boundary term evaluated on the reduced boundary of
such a portion.

Although we were motivated by the action principle in general relativity,
we find a meaningful definition of reduced boundary only in the Riemannian
sector developed according to the recipe of our Sect. $5$ (otherwise the
norm in Eq. (3.13) becomes a pseudo-norm, and it is unclear, at least
to us, what limit should be taken). At the level of ideas,
this is our contribution, which is not a direct consequence of 
previous investigations in general relativity or mathematics,
since we have not studied minimal surfaces.
The Riemannian nature of the concept we have defined is not necessarily a
failure of our research. In this respect, we recall once more the profound
result in Ref. \cite{GH1977}, where the fundamental relation between
black hole entropy and area of the event horizon was obtained from
a tree-level evaluation of the partition function in Euclidean quantum gravity.

Last but not least, our calculations in Sec. 4 show that the
abstract concepts summarized in Secs. 1-3 lead to computable results
in a gravitational background. In fact, we have addressed in Sec. 4.1, for the first time in the literature, the problem of the evaluation of the entropy of a black hole (see Eq. (\ref{corrected-entropy})) by inserting in the Euclidean action the measure-theoretic dual normal $\nu^{* \mu}$, which generalizes the usual differential-geometric picture of the normal vector field adopted in general relativity. This calculation might even imply the generalization of the concept of black hole. Indeed, in general relativity, the black is defined as each non-empty connected component of  the black hole region $\mathscr{B}$ of the spacetime $M$. This, in turn, is defined as \cite{Hawking-Ellis,Landsman2021}
\begin{equation}
\mathscr{B} = M \setminus J^{-} \left(\mathscr{J}^{+}\right),
\end{equation}
where $J^{-}\left(\mathscr{J}^{+}\right)$ denotes the causal past of  future null infinity of $M$.
On the other hand, we should correctly say that our Eq. (\ref{corrected-entropy}) deals with the calculation of the quantum entropy resulting from the reduced boundary of a finite-perimeter portion of the Euclidean Schwarzschild spacetime.

\section*{Acknowledgements}
G. E.  is grateful to Dipartimento di Fisica ``Ettore Pancini'' for hospitality and support,
and to Luigi Ambrosio, Francesco D'Andrea and Francesco Maggi for enlightening correspondence. E. B. 
is very grateful to Leo Brewin for precious correspondence and for valuable discussions regarding the 
topic of Riemann normal coordinates. Special thanks are due to Flavio Mercati for scientific 
collaboration over the whole first year of this work. E. B. dedicates this paper to 
Federica for having strongly supported him during the last year.

\begin{appendix}

\section{Borel sets and Borel functions}
\setcounter{equation}{0}

Let us denote by ${\cal P}(E)$ the set of parts of the set $E$. A family
${\cal B} \subset {\cal P}(E)$ is a Borel family if and only if it contains
the empty set and the following conditions hold with $E \in {\cal B}$:
$$
i) \; A \in {\cal B} \Longrightarrow E-A \in {\cal B},
$$
$$
ii) \; A_{i} \in {\cal B} \Longrightarrow 
\cup_{i=1}^{\infty} A_{i} \in {\cal B},
$$
$$
iii) \; A_{i} \in {\cal B} \Longrightarrow 
\cap_{i=1}^{\infty}A_{i} \in {\cal B}.
$$
For each family $S \subset {\cal P}(E)$ there exists a minimal Borel
family containing $S$ which is said to be the Borel family generated 
by $S$, and is denoted by ${\cal B}(S)$.

By definition, in a topological space, the sets belonging to the Borel
family generated by the topology are said to be Borel sets.
Thus, every set in a topological space that can be
formed from open or closed sets by means of countable union, countable intersection, and
complement, is a Borel set.
A Borel function \cite{book1} is a map $f: X \rightarrow Y$ such that $X$ and $Y$ are
topological spaces and the preimage $f^{-1}(E)$ is a Borel subset of $X$ whenever $E$ is an
open set of $Y$. One proves that this is equivalent to having $f^{-1}(E)$ as a Borel subset
of $X$ whenever $E$ is a Borel subset of $Y$. Modern developments study the role of Borel
vector fields in solving ordinary differential equations with the associated transport 
equation \cite{Gusev}.

\section{Density of points and reduced boundary}
\setcounter{equation}{0}

We here collect some definitions that are well known to expert 
research workers in geometric measure theory, but are hardly
ever met by the majority of general relativists.

Given a subset $E$ of ${\bf R}^{n}$ and a point $x$ of ${\bf R}^{n}$, if the limit \cite{book6}
\begin{equation}
\theta_{n}(E)(x) \equiv  \lim_{r \to 0^{+}}
{m_{L} (E \cap A(x,r)) \over \omega_{n}r^{n}}
\label{(B1)}
\end{equation}
exists, it is called the $n$-dimensional {\it density} of $E$ at $x$ \cite{book6}.
Given a real number $t$ in the closed interval $[0,1]$, the set of points of density $t$ 
of $E$ is defined as
\begin{equation}
E^{(t)} \equiv \left \{ x \in {\bf R}^{n}:
\theta_{n}(E)(x)=t \right \}.
\label{(B2)}
\end{equation}
A theorem ensures that every Lebesgue-measurable set is equivalent to the set of its points
of density $1$. The set $E^{1}$ is the measure-theoretic interior of $E$, while
$E^{0}$ is the measure-theoretic exterior of $E$ \cite{Ming}. 
The {\it essential boundary} is defined by
\begin{equation}
\partial^{*}E \equiv {\bf R}^{n} \backslash (E^{0} \cup E^{1}),
\label{(B3)}
\end{equation}
and is the measure-theoretic boundary \cite{Ming}  
of $E$. The various concepts of boundary are
related by the chain of inclusions \cite{Ming}
\begin{equation}
{\cal F}E \subset E^{1/2} \subset \partial^{*}E \subset \partial E.
\label{(B4)}
\end{equation}

A point $x$ belongs to the reduced boundary of a Lebesgue-measurable set $E$ if, in an open
neighbourhood of $x$, the topological boundary ${\partial E}$ of $E$ is of class $C^{1}$,
and $x$ belongs to ${\partial E}$. On the contrary, a point $x$ does not belong to the reduced
boundary of $E$ as soon as either its density differs from ${1 \over 2}$, or its density
equals ${1 \over 2}$ but the blow-up of $E$ at $x$, defined as
\begin{equation}
\Phi_{x,\rho} \equiv {(E-x)\over \rho},
\label{(B5)}
\end{equation}
is not a half-space \cite{book6}. 
The latter two conditions are not necessary and sufficient, but
help a lot in visualizing the reduced boundary of a measurable subset of ${\bf R}^{n}$. 

\end{appendix}

\end{document}